\documentclass[twocolumn,aps,prl,showpacs]{revtex4-1}
\usepackage{amsmath} 
\usepackage{amssymb} 
\usepackage{graphicx} 
\usepackage{epsfig}
\usepackage{epstopdf}
\usepackage{color}
 \usepackage{rotating}
 
\begin{document} 
\title{Pseudogap in cuprate and organic superconductors} 
\author{J. Merino$^1$ and O. Gunnarsson$^2$} 
\affiliation{$^1$ Departamento de F\'isica Te\'orica de la Materia Condensada, Universidad 
Aut\'onoma de Madrid, Madrid 28049, Spain \\
$^2$ Max-Planck-Institut f\"ur Festk\"orperforschung, Heisenbergstrasse 1, D-70569 Stuttgart, Germany} 
 
\begin{abstract}
We study the pseudogap present in cuprate and organic superconductors. We use the dynamical cluster
approximation (DCA), treating a cluster embedded in a bath. As the Coulomb interaction is increased,
cluster-bath Kondo states are destroyed and bound cluster states formed. We show that this leads 
to a pseudogap. Due to weaker coupling to the bath for the anti-nodal point, this happens 
first for this point, explaining the ${\bf k}$-dependence of the pseudogap. The pseudogap can be
understood in terms of preformed $d$-wave pairs, but it does not prove the existence of such pairs. 
\end{abstract}
\date{\today} 
\maketitle 
Experiments show a pseudogap at the Fermi energy in the normal phase
in both cuprate\cite{Timusk} and organic\cite{organics} superconductors under certain 
circumstances. For cuprates the pseudogap forms for ${\bf k}=(\pi,0)$ while for
${\bf k}=(\pi/2,\pi/2)$ the spectrum has a peak\cite{Shen}. 
It is crucial for the understanding of these systems to trace the 
origin of the pseudogap. 

Calculations using embedded cluster methods\cite{Maier}, e.g., the dynamical cluster 
approximation (DCA) reproduce a ${\bf k}$-dependent pseudogap for the Hubbard 
model\cite{Civelli,Macridin,Haule,Kyung,Ferrero,Millis,Liebsch,Sordi}. Ferrero {\it et al.}\cite{Leo}
discussed small embedded clusters in terms of a transition between a state where 
the cluster orbitals form Kondo-states with the bath and a state where the cluster forms a 
bound state and a pseudogap. However, the strong ${\bf k}$-dependence of the pseudogap in cuprates
was not discussed.  More recently, the pseudogap has been interpreted
as a momentum-selective metal-insulator transition\cite{Millis}. There has been much
work relating the pseudogap to preformed superconducting pairs\cite{Emery}, which have      
have not reached phase coherence and superconductivity at the temperature $T$ studied.
On the other hand it has been argued that the pseudogap and
superconductivity phases compete\cite{Gull}. 

The DCA treats a cluster of $N_c$ atoms embedded in a bath. Guided by the smallest 
($N_c=4$) cluster giving a pseudogap, we construct a very simple two-site two-orbital 
model. For a small Coulomb interaction $U$, the cluster orbitals form Kondo states with
the bath. As $U$ is increased, the Kondo states are destroyed and a bound state is formed
on the cluster. We show that if this state is nondegenerate, this leads to a pseudogap.
By comparing correlation functions, we find that a $N_c=8$ DCA calculation behaves
in a similar way. Due to the weak dispersion around ${\bf k}=(\pi,0)$, the coupling of 
this cluster ${\bf k}$-vector to the bath is much weaker than for ${\bf k}=(\pi/2,\pi/2)$.
Then the Kondo state for $(\pi/2,\pi/2)$ is destroyed for a larger $U$, giving
the ${\bf k}$-dependent pseudogap. We show why the pseudogap is lost for a frustrated
electron-doped cuprate. We find that the pseudogap can be interpreted in terms of $d$-wave 
superconducting pairs. The pseudogap hinders the propagation of pairs and hurts superconductivity. 

Below we study the Hubbard model
\begin{equation}\label{eq:0}
H=\sum_{ij \sigma}t_{ij}(c^{\dagger}_{i\sigma}c^{\phantom \dagger}_{j\sigma}+{\rm h.c.})
+U\sum_in_{i\uparrow}n_{i\downarrow}-\mu\sum_{i\sigma}n_{i\sigma},
\end{equation}
where $c^{\dagger}_{i\sigma}$ creates an electron on site $i$ with spin $\sigma$,
$n_{i\sigma}=c^{\dagger}_{i\sigma}c^{\phantom \dagger}_{i\sigma}$, $t_{ij}=t$ if
$i$ and $j$ are nearest neighbors,  $t'$ if they are second nearest neighbors
and zero otherwise. Here we use $t=-0.4$ eV and $t'=0$ or $-0.3t$. $U$ is a 
Coulomb integral and $\mu$ the chemical potential. 

We solve this model using DCA, and discuss the cluster in terms of its $N_c$
${\bf K}$-states. With $N_c=8$ there are both nodal [${\bf K}=(\pm \pi/2,\pm \pi/2)$] 
and antinodal [$(\pi,0), (0,\pi)$] states. For {\it isolated} clusters, the 
nodal and antinodal spectra are identical for $N_c=8$ and 16. This is due to a symmetry 
for these values of $N_c$\cite{Dagotto}. Since DCA calculations with $N_c=8$ and 16 
give a pseudogap for the antinodal point, this must then be due to the coupling to the 
bath. To test this we have switched the baths. The pseudogap then indeed appears at the 
nodal point instead of the antinodal point. An important clue is that the coupling of 
the bath to the antinodal point is weaker by a factor of three to four, due to the weaker 
dispersion of the band around this point. For a self-consistent DCA calculation,  we find 
that the bath also tends to have a pseudogap. Is then the pseudogap in the spectrum mainly
due to the pseudogap in the bath or to cluster properties? We find that performing only 
one iteration, with a fully metallic bath, the pseudogap is reproduced, although for a 
larger $U$ value. Below we then focus on the first iteration. The key issue is then
to understand the associated multi-orbital quantum impurity. We assign the pseudogap 
to the weakness of the antinodal coupling and internal electronic structure of the cluster.

The smallest embedded cluster with a pseudogap is $N_c=4$.  ${\bf K}=(0,0)$ is mainly 
occupied and ${\bf K}=(\pi,\pi)$ mainly unoccupied. We can then focus on ${\bf K}=(\pi,0)$ 
and $(0,\pi)$, which mainly determine the dynamics.  For simplicity, we first only couple 
each ${\bf K}$-state to one bath state. This leads to a two-site model with a two-fold 
orbital degeneracy.  This model has the essential features of a pseudogap and we can study 
the origin in detail. We introduce a hopping integral $V$ between each cluster orbital, labeled 
1c or 2c, and its bath site orbital, labeled 1b or 2b. On the cluster site $c$ there is 
a direct Coulomb integral $U_{\rm xx}\equiv U-\Delta U$ between spin up and down states
of the same orbital,  $U_{\rm xy}\equiv U+\Delta U$ between two different orbitals  
and an exchange integral $K$. The lowest $S_z=0$  states on the isolated $c$ site are       
\begin{eqnarray}\label{eq:1}
&&|1\mp\rangle={1\over \sqrt{2}}(c^{\dagger}_{1c\uparrow}c^{\dagger}_{1c\downarrow}\mp
                               c^{\dagger}_{2c\uparrow}c^{\dagger}_{2c\downarrow}); \hskip0.1cm E_{1\mp}=U_{\rm xx}\mp K  \\
&&|2\mp\rangle={1\over \sqrt{2}}(c^{\dagger}_{1c\uparrow}c^{\dagger}_{2c\downarrow}\mp
                               c^{\dagger}_{2c\uparrow}c^{\dagger}_{1c\downarrow}); \hskip0.1cm E_{2\mp}=U_{\rm xy}\mp K,  \nonumber
\end{eqnarray}
where $c^{\dagger}_{ic\uparrow}$ creates a spin up electron in level i on site $c$   
and $E_{i\mp}$ are the energies of the states. The state $|2-\rangle$ is a triplet and 
the other states are singlets. To simulate the lowest three states of an isolated cluster 
with $N_c=4$, we identify $1c$ with $(\pi,0)$ and $2c$ with $(0,\pi)$ and put $\Delta 
U=0.03U$ and $K=0.1U$. The lowest state in the $N_c=4$ cluster and the model cluster are 
both singlets. $U_{\rm xx}<U_{\rm xy}$ may seem surprising, but is related to the influence 
of the $(0,0)$ and $(\pi,\pi)$ orbitals, neglected in the model cluster. Alternatively, we 
can choose $U_{\rm xx}>U_{\rm xy}$. Then the lowest model cluster state is a triplet and,  
we will see, the physics is completely changed.

\begin{figure}
{\rotatebox{-90}{\resizebox{6.0cm}{!}{\includegraphics {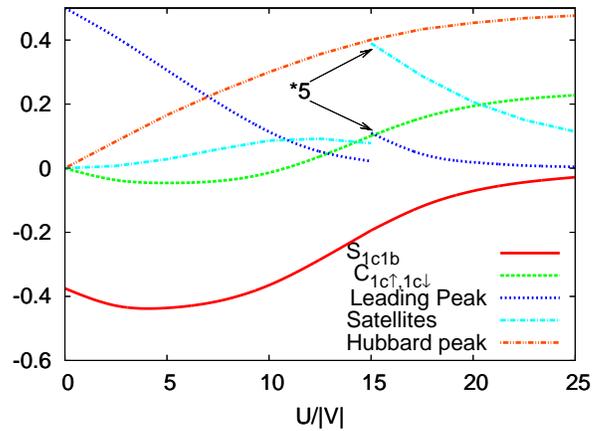}}}}
\caption{\label{fig:1}Weight of the correlation functions $S_{\rm 1c1b}$ and 
$n_{1c \uparrow}n_{1a \downarrow}$ as well as spectral weights for the leading peak,
satellites forming the peaks surrounding the the pseudogap and the Hubbard peak for the two-level two-orbital model.
}
\end{figure}

For the two-level, two-orbital model we have calculated the correlation function
$S_{\rm 1c1b}=\langle {\bf S}_{1c}\cdot {\bf S}_{1b}\rangle$, between the spins in orbitals 1c and 1b, 
on the cluster and in the bath. We have $-3/4\le S_{\rm 1c1b}$. This is shown in Fig.~\ref{fig:1}
as a function of $U$ (full curve). For moderate values of $U$ there is a strong 
negative correlation, which increases with $U$.  The correlation function 
$C_{{\rm 1c}\uparrow {\rm 1c}\downarrow}=\langle n_{\rm 1c\uparrow } n_{\rm 1c \downarrow}\rangle-1/4$ 
is also shown. It   becomes negative as U is increased, implying a reduced double 
occupancy of orbital 1c and a spin 1/2 state starts to form on 1c. Both these results
are consistent with levels 1c and 1b forming a Kondo-like state. Further increase of $U$, 
however, leads to a rapid reduction of $|S_{\rm 1c1b}|$ and $C_{{\rm 1c}\uparrow {\rm 1c}\downarrow}$
grows positive. The Kondo singlets are then broken, and instead a bound state is formed 
on the cluster. Increasing $U$ reduces the Kondo energy, while the  gain from correlating 
the electrons on the cluster increases. In the lowest cluster state  
$C_{{\rm 1c}\uparrow {\rm 1c}\downarrow}$  takes the value 0.25, which is approached in Fig.~\ref{fig:1}.

Fig.~\ref{fig:1} also shows weights of the spectral function on the photoemission side
for the two-orbital, two-site model. 
For $U=0$ all weight is in the leading peak.  As $U$ is increased, the weight of the Hubbard peak 
increases, corresponding to final state with primarily one electron less on the cluster site. 
At the same time three peaks close to the leading peak get more weight, summed up in the curve 
``Satellites''. These peaks correspond to final states with substantial weights in excited neutral 
cluster states, while the leading peak corresponds to a final state with a large weight
of the lowest neutral state. For $U/|V|\sim 11$ the satellites acquire more weight than the leading peak and 
a pseudogap starts to develop. This happens when the Kondo-like effect starts to be lost 
and a bound state on the cluster starts to form.         

\begin{figure}
{\rotatebox{-0}{\resizebox{6.0cm}{!}{\includegraphics {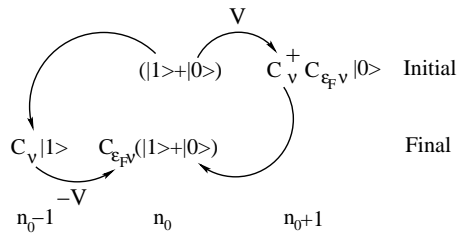}}}}
\caption{\label{fig:2}Initial and final states for a process
removing an electron with quantum number $\nu$. 
There is a negative interference 
between two channels for reaching the final state.
}
\end{figure}

To address the pseudogap, we study a general cluster with $U>>|V|$. Assume an integer number of 
electrons per cluster, $n_0$, and project out the piece, $|\Phi_{n_0}\rangle$, of the ground-state, 
$|\Phi\rangle$, that has exactly $n_0$ electrons. We study   photoemission processes, removing an 
electron with the quantum number $\nu$ and energy $\varepsilon$. We split $|\Phi_{n_0}\rangle$ as             
\begin{equation}\label{eq:2}
|\Phi_{n_0}\rangle=|0\rangle+|1\rangle,
\end{equation}
where $c_{\nu}|0\rangle=0$ and  $c^{\dagger}_{\nu}c^{\phantom \dagger}_{\nu}|1\rangle=|1\rangle$.
Let $H_0=\sum_{\nu\varepsilon} (V_{\nu\varepsilon}c^{\dagger}_{\nu}c^{\phantom \dagger}
_{\nu \varepsilon}+{\rm h.c.})$ give the hopping into the cluster.
We then  write the part of the ground-state corresponding to $n_0+1$ electrons as
\begin{equation}\label{eq:3}
|\Phi_{n_0+1}\rangle=-\sum_{\nu^{'}}\sum_{\varepsilon}{V_{\nu^{'} \varepsilon}\over \Delta E_+-\varepsilon}
c^{\dagger}_{\nu^{'}}c^{\phantom \dagger}_{\nu^{'}\varepsilon}|\Phi_{n_0}\rangle,
\end{equation}
where we have approximated the energy difference between cluster states with one extra electron (hole) and 
the lowest neutral state as $\Delta E_+$ ($\Delta E_-$).
If the lowest state of the isolated cluster is nondegenerate, this is (nondegenerate) perturbation theory.
However, if the lowest state is degenerate, e.g., for the Kondo problem, $|0\rangle$ and $|1\rangle$ 
contain much more information than can be obtained from perturbation theory, and the treatment goes 
beyond perturbation theory. For an infinite system, we construct a state $|\nu \varepsilon_F  \rangle$ with
a hole in the bath at the Fermi energy $\varepsilon_F$. We form $c_{\nu\varepsilon_F}|\Phi_{n_0}\rangle$ and allow 
hopping to states with $n_0-1$ electrons on the cluster. The corresponding amplitude is then
\begin{eqnarray}\label{eq:4}
&&\langle \nu \varepsilon_F |c_{\nu}|\Phi\rangle=
\sum_{\varepsilon}V_{\nu\varepsilon}({\langle 1| c_{\nu\varepsilon_F}^{\dagger}c_{\nu\varepsilon}^{\phantom \dagger}|1\rangle \over \Delta E_-+\varepsilon}
-{\langle 0| c_{\nu\varepsilon_F}^{\dagger}c_{\nu\varepsilon}^{\phantom \dagger}|0\rangle \over \Delta E_+-\varepsilon})
\nonumber\\
&&+\sum_{\nu^{'}(\ne\nu)\varepsilon} V_{\nu^{'} \varepsilon}({\langle 0|c^{\dagger}_{\nu^{'}}c^{\phantom \dagger}_{\nu^{\phantom '}}c^{\dagger}_{\nu^{\phantom '}\varepsilon_F }c^{\phantom \dagger}_{\nu^{'}\varepsilon }|1\rangle\over \Delta E_-+\varepsilon} 
+{\langle 0|c^{\dagger}_{\nu^{'}}c^{\phantom \dagger}_{\nu^{\phantom '}} c^{\dagger}_{\nu^{\phantom '}\varepsilon_F }c^{\phantom \dagger}_{\nu^{'}\varepsilon }|1\rangle\over \Delta E_+-\varepsilon}). \nonumber          
\end{eqnarray}
If the isolated cluster ground-state is nondegenerate and $\Delta E_{\pm}$ are large, 
the number of holes in the bath for the states $|0\rangle$ and $|1\rangle$ is small.     
Then the third and fourth terms are also small, while the expectation values in 
the first and second terms are large and tend to cancel. If the orbital $\nu$ is 
half-filled the cancellation becomes perfect. This cancellation is illustrated 
schematically in Fig.~\ref{fig:2}. For, e.g., $U=1.0$ and $V=-0.02$ in the model above, 
the four terms are -0.0117, 0.0121, 0.0020, 0.0020. This illustrates the strong cancellation 
between the first two terms and the small magnitude of the next two terms. 

If the lowest cluster state is degenerate, the situation is quite different.        
The system may then form a Kondo-like state even for large $U$. The first
term in the amplitude above is then smaller and does not cancel the second term. 
Furthermore, the ``spin-flip'' terms three and four can be substantial.                       
As a result, a pseudogap may not develop. In the model above, a negative      
$\Delta U$ gives a triplet ground-state for the model cluster. In the limit of 
a large $U$, there are then triplet states on both the c and b site, coupling to
a total singlet. The leading peak has more weight than each of the neighboring 
 satellite for the $U$ range shown in Fig.~\ref{fig:1}, and a pseudogap does not 
form. For an infinite system, these results can also be discussed in terms 
of the phase shift at the Fermi energy\cite{Leo}. 

\begin{figure}
{\rotatebox{-90}{\resizebox{6.0cm}{!}{\includegraphics {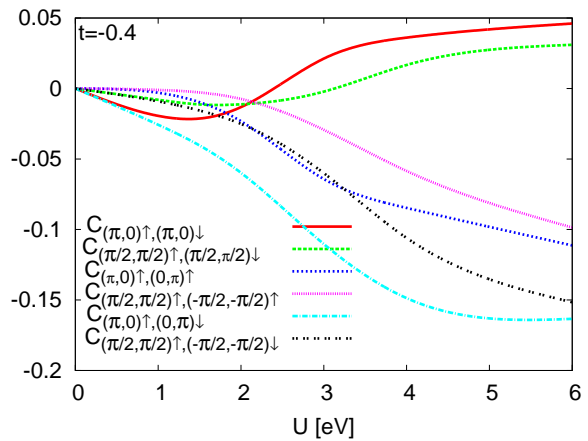}}}}
\caption{\label{fig:3}Correlation functions as a function of $U$ for an embedded
eight site cluster in the first iteration.The parameters are $t=-0.4$ eV and $T=290$
K.
}
\end{figure}

We now study $N_c=8$ at half-filling. Fig.~\ref{fig:3} shows correlation functions of the type
$C_{(\pi,0) \uparrow,(\pi,0) \downarrow}=\langle n_{(\pi,0)\uparrow}n_{(\pi,0)\downarrow}\rangle
-\langle n_{(\pi,0)\uparrow}\rangle \langle n_{(\pi,0)\downarrow}\rangle$.
This function behaves in a very similar way as $C_{{\rm 1c}\uparrow {\rm 1c}\downarrow}$ in Fig.~\ref{fig:1}.
As $U$ is increased it first turns negative, indicating the formation of a Kondo-like
state. Then it turns positive, indicating the formation of a 
bound state on the cluster in the $(\pi,0), (0,\pi)$-space similar as for the 
two-site two-orbital model. This is also supported by the strong negative
correlation between $(\pi,0)$ and $(0,\pi)$ as is also found for $|1-\rangle$ in Eq.~(\ref{eq:1}).
At the same time a pseudogap develops in the $(\pi,0)$ spectrum, as in Fig.~\ref{fig:1}. 
$C_{(\pi/2,\pi/2) \uparrow,(\pi/2,\pi/2) \downarrow}$ behaves in a similar way,   
but it stays Kondo-like up to larger values of $U$ before the $(\pm\pi/2,\pm \pi/2)$-space is also 
used to form a bound state on the cluster. Therefore the $(\pm\pi/2,\pm \pi/2)$ pseudogap forms for 
larger $U$. The reason is that the coupling to the bath is  a factor of three to four stronger
for $(\pi/2,\pi/2)$ and it is favorable to keep the Kondo-like state
up to a larger $U$.

To further support this, we have performed exact diagonalization (ED) calculations 
for $N_c=8$ and one bath level  at the chemical potential with the coupling
$V({\bf K})$ to the cluster ${\bf K}$-state. We study the correlation 
$\langle S^zs^z\rangle$ between spins on the cluster and the bath and the hopping 
energy $\langle V \rangle$ between the cluster and the bath for a certain ${\bf K}$. 
Table~\ref{table:1} shows that both quantities are much larger for $(\pi/2,\pi/2)$ 
and that both decrease rapidly with $U$. 

\begin{table}
\caption{Bath-cluster spin correlations and hybridization energies from ED calculations.
The parameters are $t=-0.4$ eV, $V(\pi,0)=0.08$ eV and $V(\pi/2,\pi/2)= 0.21$ eV.  }
\label{table:1}
\begin{tabular}{lllllll}
\hline
\hline
  &   &    &  \multicolumn{2}{c}{$\langle S^zs^z\rangle$} & \multicolumn{2}{c}{$\langle V\rangle$}  \\
$U$ & $\mu$ & n  &  $(\pi,0)$ & $({\pi\over 2},{\pi\over 2})$ &   $(\pi,0)$  &  $({\pi\over 2},{\pi\over 2}) $   \\
\hline
2.5 & 1.25 &  1.0  &   -.0020     &   -.0051 & -.021   &  -.097\\
3.2 &  1.6 &  1.0   &   -.0012  &  -.0030     &-.014   &    -.066       \\
\hline
\end{tabular}
\end{table}

The energy gain when all electrons of the isolated cluster
correlate is much larger than the sum of the energies gained
when correlation is only allowed in the $(\pi,0), (0,\pi)$- or $(\pm\pi/2,\pm\pi/2)$-spaces.
This favors a simultaneous switch of both spaces from the Kondo-like states to
a correlated cluster state. However, the difference in the coupling to the bath
is so large that the switch, nevertheless, happens for different $U$ 
in Fig.~\ref{fig:3}. This can be different for the frustrated case.

\begin{table}
\caption{Bath-cluster spin correlations and hybridization energies from ED calculations on $N_c=8$
frustrated clusters with $U=2.5$ eV, $t=-0.4$, $t'=-0.3t$, $V(\pi,0)=0.08$ eV  and $V(\pi/2,\pi/2)= 0.21$ eV.
We also show $J({\bf K})$ [Eq.~(\ref{eq:5})].}
\label{table:2}
\begin{tabular}{llllllll}
\hline
\hline
      &    &  \multicolumn{2}{c}{$\langle S^zs^z\rangle$} & \multicolumn{2}{c}{$\langle V\rangle$} & \multicolumn{2}{c}{$J/V^2$} \\
$\mu$ & n  &  $(\pi,0)$ & $({\pi\over 2},{\pi\over 2})$ &   $(\pi,0)$  &  $({\pi\over 2},{\pi\over 2}) $ & $(\pi,0)$  & $({\pi\over 2},{\pi\over 2})$\\
                \hline
                0.85 &  0.96   &  -.0042    &    -.0068    &  -.036  &    -.12  & 1.61 & 1.78  \\
                        1.05  &   0.98     &  -.0038   &    -.0063  & -.035   & -.11 & 1.69 & 1.64 \\
                        1.25 &  1.0  &   -.0048       &    -.0065 & -.043   &  -.11 & 1.88 & 1.60 \\
                        1.35  &  1.02    &  -.0064     &      -.0063   &  -.047  &  -.12  & 2.04 & 1.61 \\
                        1.45  &   1.04     &  -.0084   &    -.0071  & -.055   & -.13 & 2.27 & 1.64 \\
\hline
\end{tabular}
\end{table}

We now consider the frustrated case, $t'=-0.3t$, for different fillings and $U=8|t|$
in the first iteration.        
For moderate hole doping the  $N_c=8$ calculation shows a clear pseudogap.
On the other hand, for electron doping we find no pseudogap. We first study 
this using ED as above. The coupling to the bath is kept fixed, and $n$ is 
varied by varying  $\mu$.  Table~\ref{table:2} shows how both the cluster-bath spin 
correlation and hopping energy increase for $(\pi,0)$ as we go from hole- to electron-doping,
while the change is small for $(\pi/2,\pi/2)$. To understand this,
we study the $J({\bf K})$ obtained from a Schrieffer-Wolf transformation
\begin{equation}\label{eq:5}
J({\bf K})=|V({\bf K})|^2{\Big(} {1 \over \epsilon({\bf K})+ U-\mu}+{1\over \mu - \epsilon({\bf K}) }{\Big)}. 
\end{equation}
Due to the shift of $\mu$ with $n$, $J({\bf K})$ increases for $(\pi,0)$, 
while the levels for $(\pi/2,\pi/2)$ are located in such a way that the changes
of $J({\bf K})$ are relatively small.

In the ED calculation we assumed a doping independent coupling.
The coupling is related to Im $1/G_0({\bf K},\omega_n)$, where
$G_0$ is the Green's function of the bath. For $t'=-0.3t$ and  moderate hole doping,
both Re $G_0[(\pi,0),\omega_n]$ and Im $G_0[(\pi,0),\omega_n]$ are large for
small imaginary frequencies $\omega_n$. Then Im $1/G_0[(\pi,0),\omega_n]$ is
small and the coupling relatively weak. As $n$ is increased, $\mu$ shifts so that 
for the electron doped system both the real and imaginary parts of $G_0$ are smaller 
and the coupling larger. This increase and the increase of $J({\bf K})$ with $n$ 
favor a Kondo effect for $(\pi,0)$ for the electron-doped system. The Kondo effect then 
tends to be lost simultaneously in the $(\pi,0)$- and $(\pm\pi/2,\pm\pi/2)$-spaces. 
This result depends crucially on the sign of $t'/t$.  For the organic superconductors, 
with a triangular lattice and at half-filling, we find that the pseudogap goes away at 
$t'\sim 0.6t$, in agreement with experiment\cite{organics}, since $J(\pi,0)$ has a minimum at
$t'=0$ for half-filling. An additional effect is that  when the system is doped,
the ground-state has a substantial mixture of a cluster state with one electron more 
or less. This state is degenerate, which is unfavorable for a pseudogap and in agreement 
with the lack of a pseudogap for large doping.  

To understand the character of the pseudogap, we have calculated 
$P_d=\sum_{ij}\langle \Delta_i\Delta_j^{\dagger}\rangle/N_c$, where $\Delta_i$ is the 
operator for $d$-wave superconductivity. A factorized term was subtracted so that 
$P_d=0$ for $U=0$. For the ground-state  of the isolated cluster $P_d$ is rather large 
(see Table~\ref{table:3}).  This state dominates the ground-state of the embedded 
cluster for relatively large $U$. The peaks surrounding the pseudogap correspond 
primarily to final states with a large weight of excited neutral states with $S=0$ or 1,
for which $P_d$ is substantially smaller. We can then think of the pseudogap 
as corresponding to the break up of a d-wave singlet pair. 

This relation of the pseudogap to $d$-wave pairs does not contradict the 
observation that the pseudogap and superconductivity compete\cite{Gull}.
The occurrence of superconductivity is determined
by the Bethe-Salpeter equation, and depends on the matrix $(1-\bar \chi^0 \Gamma_c)$ becoming
singular\cite{Maier}. Here  $\Gamma_c$ is the two-particle irreducible vertex,    
representing the interaction, and $\bar \chi^0$ consists of products of two dressed
Green's functions, describing propagation of pairs. $\bar \chi^0$ is closely related 
to the density of states at the Fermi energy. As $U$ is increased, $\Gamma_c$ grows and 
$\bar \chi^0$ is reduced. This reduction becomes very rapid as a pseudogap forms\cite{Jarrell}.
This hurts the onset of superconductivity. A similar competition 
between $\bar \chi^0$ and $\Gamma_c$ was observed for alkali-doped fullerides\cite{c60}.

\begin{figure}
{\rotatebox{-90}{\resizebox{6.0cm}{!}{\includegraphics {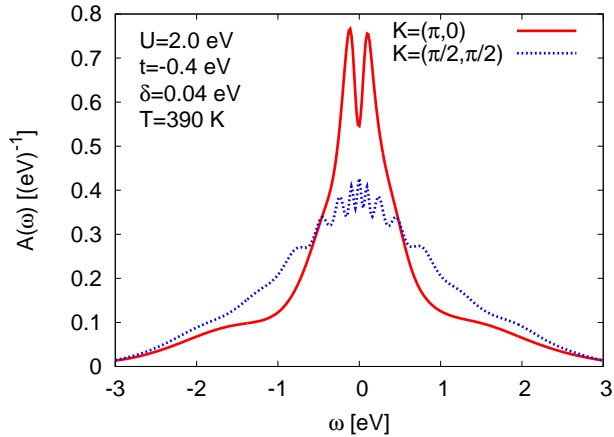}}}}
\caption{\label{fig:4}Spectrum averaged over ${\bf k}$ around ${\bf K}=(\pi,0)$ and $(\pi/2,\pi/2)$ 
when an AF field $\delta$ is applied to the cluster.
}
\end{figure}

From the discussion below Eq.~(\ref{eq:3}) we may expect to see a pseudogap
if the lowest state of the cluster is nondegenerate, independently of $P_d$.
To test this, we have applied an antiferromagnetic (AF) field  $\sum_i\delta_i(n_{i\uparrow}-n_{i \downarrow})$,   
where $\delta_i=\delta$ on one sublattice and $-\delta$ on the other. 
This term is only applied to the cluster and not to the bath. One iteration is performed.
This term favors an AF state on the cluster and reduces $P_d$. Fig.~\ref{fig:4}
shows that there is, nevertheless, a pseudogap for ${\bf K}=(\pi,0)$.

\begin{table}
\caption{\label{table:3}Energy $E$, spin $S$, degeneracy Deg. and $d$-wave superconductivity
correlation $P_d$ for low-lying states with $S_z=0$ for $N_c=8$.
Deg. refers to the degeneracy of $S_z=0$ states.  The parameters are $U=3$, $t=-0.324$ eV and $t'=0$.
}
\begin{tabular}{cccccc}
\hline
\hline
$E$ & -1.116 & -1.062 & -0.941 & -0.897 & -0.764 \\
$S$ & 0 & 1 & 2& 1 & 0 \\
Deg. & 1 & 1 & 1 & 6 & 9 \\
$P_d$ & 0.37 & 0.28 & 0.10 & 0.15 & 0.10\\
\hline
\end{tabular}
\end{table}

In conclusion we have provided and analyzed a very simple model, showing pseudogap 
formation. As $U$ is increased, the system switches from a case where the cluster 
orbitals form a Kondo state with the bath to a case where a nondegenerate bound state is formed 
on the cluster. This leads to a pseudogap. We showed that this analysis also applies to 
larger clusters. It is crucial that the coupling to the bath is much weaker for ${\bf K}=(\pi,0)$
than for $(\pi/2,\pi/2)$, which leads to the strong ${\bf K}$-dependence. We have found that 
the pseudogap is lost for the electron doped system, due to an increase in the effective $J$
and the cluster-bath coupling for ${\bf K}=(\pi,0)$. The pseudogap can be interpreted in terms
of $d$-wave superconducting pairs. It strongly reduces $\bar \chi^0$, describing how the propagation of 
pairs is hindered, leading to a competion with superconductivity.

JM acknowledges financial support from MINECO (MAT2011-22491) and hospitality at Max-Planck-Institut 
f\"ur Festk\"orperforschung in Stuttgart during his stay there.

\end{document}